\documentclass[aps,amsmath,amssymb,reprint,groupedaddress, showpacs]{revtex4-1}

\usepackage{graphicx}% Include figure files
\usepackage{dcolumn}% Align table columns on decimal point
\usepackage{bm}% bold math
\usepackage[greek, ngerman, english]{babel}

\usepackage[ separate-uncertainty  = true, 
   uncertainty-separator =  {\,}, 
   mode = text,
   exponent-product = \cdot,
   output-product = \cdot,
   output-decimal-marker ={.}, 
   multi-part-units      =  single, 
   range-phrase          = {--}]{siunitx}

\begin{document}

%\preprint{}

\title{Azimuthal asymmetries in SIDIS di-hadron muoproduction off longitudinally polarized protons at COMPASS}



%\author{Bakur Parsamyan}
%\affiliation{University of Turin \& INFN}
%\email[]{bakur.parsamyan@cern.ch}

\author{Stefan Sirtl} \thanks{The author acknowledges financial support by the German Bundesministerium f\"ur Bildung und Forschung (BMBF)}
\affiliation{Physics Department, Albert-Ludwigs-University, Freiburg 79104, Germany}
\email[]{mail@stefansirtl.de}
%\thanks{}

\collaboration{On Behalf of the COMPASS Collaboration}


\date{\today}

\begin{abstract}
In this review a first comprehensive study of azimuthal asymmetries in semi-inclusive deep inelastic muoproduction of hadron pairs off longitudinally polarized protons at COMPASS is presented. The study is based on data taken in 2007 and 2011, obtained by impinging a high-energetic $\mu^+$ beam of $\SI{160}{GeV/\it{c}}$, respectively $\SI{200}{GeV/\it{c}}$, momentum on a solid ammonia target. The discussion is focused on both leading and subleading longitudinal target-spin-dependent asymmetries arising in the di-hadron SIDIS cross section, addressing the role of spin-orbit couplings and quark-gluon correlations in the framework of collinear or transverse momentum dependent factorization.
\end{abstract}

%\keywords{}

\maketitle


\section{Introduction}
Azimuthal cross section asymmetries in semi-inclusive deep inelastic scattering (SIDIS) of polarized leptons off polarized nucleons are key observables to investigate the spin dependent substructure of the nucleon. Assuming factorization, azimuthal asymmetries can be theoretically connected to combinations of parton distribution functions (PDFs) and fragmentation functions (FFs), encoding information about the partonic substructure of the nucleon and the fragmentation mechanism, respectively.
In this review we present a first study of azimuthal target-spin-dependent asymmetries, arising in the di-hadron SIDIS cross section, measured on longitudinally polarized protons at COMPASS. This includes the measurement of a set of nine azimuthal asymmetries, appearing in a transverse-momentum-dependent (TMD) approach at leading-twist, which can be consequently related to spin-orbit couplings. In particular, these asymmetries are sensitive to the TMD PDFs $g_{1L}$ and $h_{1L}$, describing the helicity distribution, respectively the distribution of transversely polarized quarks in a longitudinally polarized proton. \\
\par
Moreover, also azimuthal modulations at subleading-twist are considered within this work, which survive the integration over quark transverse momenta. Measuring respective collinear cross section asymmetries at subleading-twist can provide new understanding of so far unresolved quark-gluon correlation mechanisms. In particular, such results can help to access the yet unknown collinear PDFs $h_L$ and $e_L$. Our results are characterized by an unprecedented precision, covering a wide kinematic range. For more comprehensive information on this analysis the reader may be referred to Ref. \cite{Sirtl:Phd}. A similar study has been already presented by the CLAS collaboration, focused on collinear asymmetries\cite{Pereira2014}.  

%The focus in this note is set on the latter, namely the single spin asymmetry $A_{UL}^{\sin(\phi_R)}$ and the double spin asymmetry $A_{LL}^{\cos(\phi_R)}$. Being of subleading twist, measuring these observables can provide new understanding of so far unresolved quark-gluon correlation mechanisms. In particular, results can help to access the unknown PDFs $h_L(x)$ and $e_L(x)$. Additionally, nine azimuthal amplitudes with a TMD-interpretation at leading twist are presented in short. Results are extracted from data taken at COMPASS in the years 2007 and 2011 by scattering a polarized $\mu^+$ beam off a longitudinally polarized $NH_3$ target, which allows for a simultaneous measurement of both proton spin states, either parallel or antiparallel with respect to the beam direction.  

\section{Theoretical Framework}
\label{sec:Theo}

This work considers the di-hadron SIDIS process
\begin{equation}
\mu(l)+p(P) \rightarrow \mu(l')+h_{1}(P_{1})+h_{2}(P_{2})+X,
\label{Eq:Process}
\end{equation}
where a beam muon $\mu$ probes a target proton $p$ with mass $M$ via the exchange of a virtual photon. The corresponding four-momenta are given in parenthesis in the above formula. The struck quark subsequently fragments into two unpolarized final state hadrons $h_1$ and $h_2$ and any, not necessarily detected, rest $X$ in the final state. In particular, this work considers SIDIS of longitudinally polarized muons off longitudinally polarized protons, producing hadron pairs of opposite charge. In order to keep orientations well defined, $h_{1}$ is defined to be the positively and $h_{2}$ the negatively charged hadron.\\
\par

\begin{figure}[t]
\includegraphics[width=0.5\textwidth]{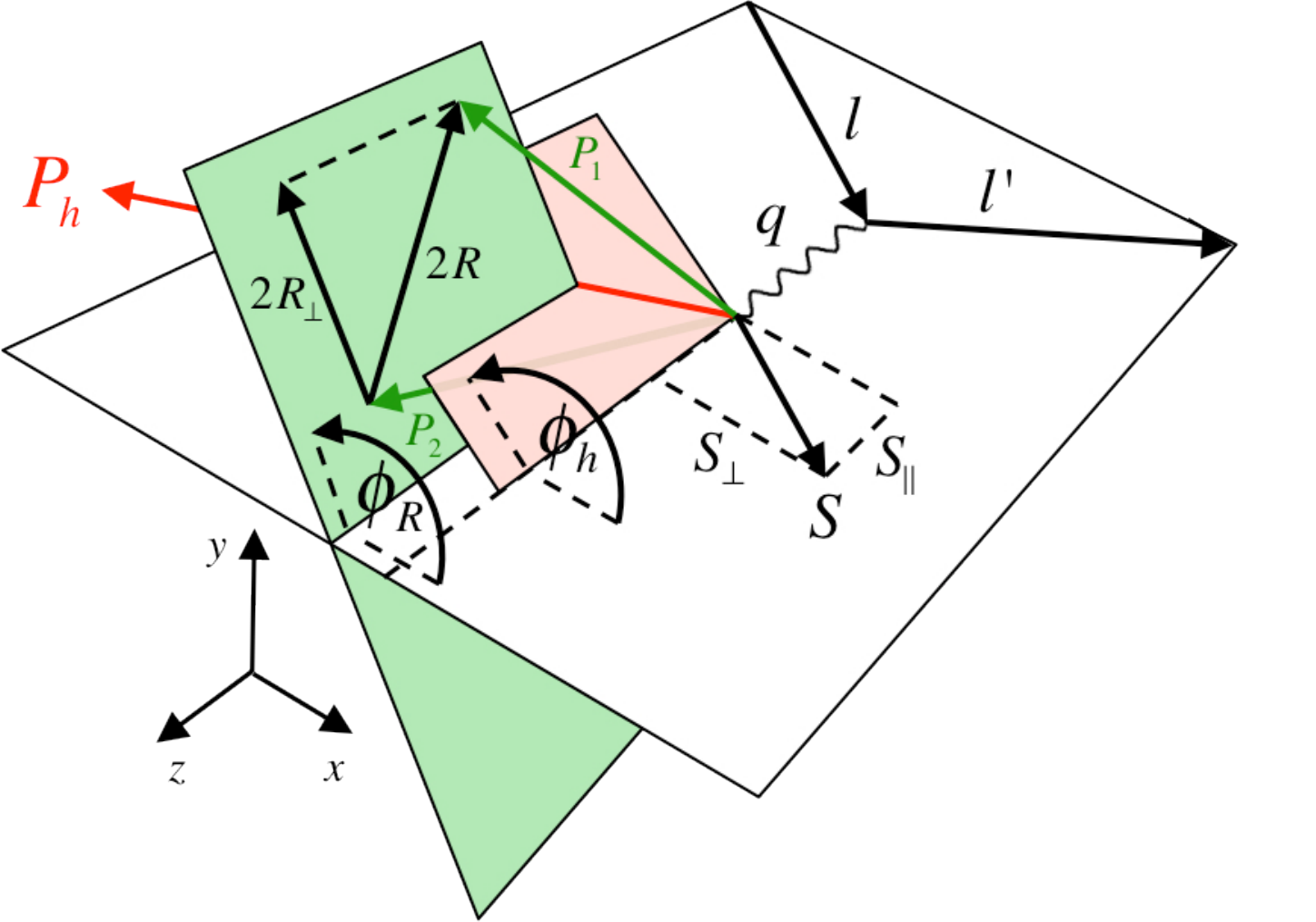}
\caption{Sketch of the considered two di-hadron SIDIS process, including the relevant azimuthal angles. The nucleon is assumed to be longitudinally polarized either along or against the direction of the incoming lepton.}
\label{fig:kinematics}
\end{figure}

The di-hadron cross-section is modulated in two azimuthal angles, $\phi_h$ and $\phi_R$ \citep{Bacchetta:2002ux, Gliske:2014aa, Bacchetta2hLT}. As sketched in Fig. \ref{fig:kinematics}  they are both enclosed by the scattering plane, spread by the incoming lepton and the virtual photon direction, and a hadronic plane, spread by the virtual photon direction and either
\begin{equation}
P_{h} = P_{1}+P_{2} \hspace{1cm} \text{or} \hspace{1cm} R = \frac{1}{2}\left(P_{1}-P_{2}\right).
\end{equation}
The azimuthal angles can be consequently calculated via
\begin{eqnarray}
\phi_{h} &=& \frac{\left( \bm{q} \times \bm{l}\right)\cdot \bm{P}_{h}}{\left|\left( \bm{q} \times \bm{l}\right)\cdot \bm{P}_{h}\right|}\arccos \left(\frac{\left( \bm{q} \times \bm{l}\right)\cdot \left( \bm{q} \times \bm{P}_{h}\right)}{\left| \bm{q} \times \bm{l}\right| \cdot \left| \bm{q} \times \bm{P}_{h}\right|}\right)\\[7pt]
\phi_{R} &=& \frac{\left( \bm{q} \times \bm{l}\right)\cdot \bm{R}_{\perp}}{\left|\left( \bm{q} \times \bm{l}\right)\cdot \bm{R}_{\perp}\right|}\arccos \left(\frac{\left( \bm{q} \times \bm{l}\right)\cdot \left( \bm{q} \times \bm{R}_{\perp}\right)}{\left| \bm{q} \times \bm{l}\right| \cdot \left| \bm{q} \times \bm{R}_{\perp}\right|}\right).
\end{eqnarray}

where the bold variables indicate corresponding momenta. Here, $\bm{R}_{\perp}$ describes the transverse component of $\bm{R}$ with respect to the virtual photon. It is calculated from
\begin{equation}
\bm{R}_{\perp}=\frac{z_{2}\bm{P}_{1\perp}-z_{1}\bm{P}_{2\perp}}{z_{1}+z_{2}}
\end{equation}
in order to ensure the invariance of $\phi_{R}$ against boosts in the direction of the virtual photon, where $z_{1/2} = E_{1/2}/\nu$ is the energy fraction of a hadron with respect to the virtual photon energy. This definition of $\bm R_\perp$ coincides with the general one up to corrections of order $1/Q^{2}$ \cite{Gliske:2014aa, KotzinianR}, where $Q^2=-(l-l')^2$.\\
\par
Asymmetries are defined as ratios of structure functions
\begin{equation}
A_{XY}^{m(\phi_h, \phi_R)}=\frac{F_{XY}^{m(\phi_h, \phi_R)}}{F_{UU,T}+\varepsilon F_{UU,L}},
\label{eq:Asy_def}
\end{equation}
where the subscripts indicate the polarization of the beam (X) and the target (Y), here either unpolarized (U) or longitudinally polarized (L). The third subscript refers to longitudinally (L) or transversely (T) polarized virtual photons. The ratio of the corresponding photon fluxes is given by
\begin{equation}
\varepsilon = \frac{1-y-\frac{1}{4}\gamma^{2}y^{2}}{1-y+\frac{1}{2}y^{2}+\frac{1}{4}\gamma^{2}y^{2}},
\end{equation}
where $y=\frac{\nu}{E}$ is the fractional energy of the virtual photon and $\gamma = \frac{2Mx}{Q}$.
The superscript $m(\phi_h, \phi_R)$ in Eq.(\ref{eq:Asy_def}) indicates the respective azimuthal modulation. \\
\par
In a TMD approach, i.e. when taking into account transverse momenta of quarks $p_T$, a set of seven azimuthal single spin asymmetries (SSAs) and two double spin asymmetries (DSAs) can be measured at leading twist. They are sensitive to $p_T$-dependent convolutions of TMD PDFs, in particular the helicity distribution $g_{1L}(x,p_T)$ or the still unknown Boer-Mulders function $h_{1L}(x,p_T)$, coming with FFs. Detailed formulas can be found in Ref. \cite{Sirtl:Phd}.\\
\par
Considering the di-hadron cross-section in a collinear approach, two longitudinal target spin asymmetries arise at subleading twist: 

 \begin{widetext}
 \begin{align}
 A_{UL}^{\sin(\phi_R)} =\:&-\frac{M}{Q}\frac{|\bm R|}{M_h}\:\frac{\sum_q{e_q^2}\left[xh_L^q(x)H_1^{\angle q,sp}(z,M_h^2)+\frac{M_h}{Mz}g_1^q(x)\tilde{G}^{\angle q,sp}(z,M_h^2)\right]}{\sum_q{e_q^2 f_1^q(x)D_{1}^{q,ss+pp}(z,M_h^2)}}\\[14pt]
 %A_{LL}^{\cos(\phi_R)} =-\frac{M}{Q}\frac{|\bm R_T|}{M_h} \:\frac{\sum_q{e_q^2 \frac{M_h}{Mz}g_1^q(x)\tilde{D}^{\angle q,sp}(z,M_h^2)}}{\sum_q{e_q^2 f_1^q(x)D_{1}^{q,ss+pp}(z,M_h^2)}}.
 A_{LL}^{\cos(\phi_R)} =\:&\hspace{0.5cm}\frac{M}{Q}\frac{|\bm R|}{M_h} \:\frac{\sum_q{e_q^2 \left[x e_L^q (x) H_1^{\angle q,sp}(z,M_h^2)-\frac{M_h}{Mz}g_1^q(x)\tilde{D}^{\angle q,sp}(z,M_h^2)\right]}}{\sum_q{e_q^2 f_1^q(x)D_{1}^{q,ss+pp}(z,M_h^2)}}.
 \end{align}
 \end{widetext}
 
Their interpretation in the framework of the parton model involve flavor sums of simple products of PDFs and FFs, in particular interference FFs ($\angle$), whereas the superscripts $s$ and $p$ indicate the contributing partial wave characteristics. The electric charge of a particular flavor $q$ is denoted by $e_q$. Assuming Wandzura-Wilzcek approximation, the genuine twist-3 terms marked with a tilde can be neglected, leaving a pure sensitivity to the respective leading products. Measuring these asymmetries, and including recent results for the interference FF $H_1^{\angle}$ from BELLE \cite{BelleIFFH1}, hence provides a clean way to access the still unknown twist-3 PDF $h_L(x)$, respectively $e_L(x)$. The first can be interpreted as the distribution of transversely polarized quarks in a nucleon with longitudinal spin orientation, which is among the missing puzzle pieces to complete the one-dimensional picture of the proton at subleading twist. Assuming the gauge-link to be the only source of t-odd behavour, the PDF $e_L(x)$ and with it the respective asymmetry should vanish. Measuring these asymmetries can hence provide further insight into $Q$-suppressed spin dependent mechanisms and serve to corroborate common theoretical assumptions.

\section{Data Analysis}

This work comprises the analysis of combined data, obtained by scattering naturally polarized $\mu^+$ with a nominal momentum of \SI{160}{GeV/\it{c}} during a dedicated data taking in 2007, respectively of \SI{200}{GeV/\it{c}} in 2011, off a longitudinally polarized solid state $\text{NH}_3$ target. A priori the $Q^2$-evolution and the kinematic dependences of the considered asymmetries are unknown. Still, from general considerations, these kind of effects are expected to be small or negligible within experimental accuracy. Hence, we find it reasonable to merge both data sets, although different beam energies were used.\\
\par
The standard COMPASS DIS cuts were applied. In particular was the four-momentum transfer limited to  $Q^{2} > \SI{1}{(GeV/\it{c}\normalfont)^{2}}$, the fractional energy transfer of the muon set to $ 0.1 < y < 0.9$ and the invariant mass of the hadronic system required to be $W > \SI{5}{GeV/\it{c}^{2}}$. To match COMPASS kinematics, the Bjorken variable was limited to $ 0.0025 < x < 0.7$. Per selected event, all possible combinations of hadron pairs were included in the analysis. The fractional energy for each hadron was required to be $z_{1/2} > 0.1$ and the Feynman variable $x_{\text{F},1/2} > 0.1$. To further exclude exclusive events from the sample, the missing energy 
\begin{equation}
E_{\text{miss}}=\frac{\left(P+q-P_h\right)^2-q^2}{2M}=\frac{M_X^2-M^2}{2M},
\end{equation}
was required to fulfill $E_{\text{miss}} > \SI{3}{GeV}$. Here, $M$ and $M_X$ stand for the mass of the proton, respectively the mass of the undetected recoiling system. Finally, a cut $R_{T} > 0.07$ was applied, to ensure the well-definition of the corresponding hadronic plane, hence the angle $\phi_R$.\\
 \par
A further remark should be given concerning the polarization of the target. Since it is practically polarized along beam direction, there enters a transverse spin contribution when considering the  frame where the z-axis points along the direction of the virtual photon. In this analysis, this contribution of transverse polarization components along the photon axis is neglected due to its strong suppression in COMPASS kinematics.\\
\par
All azimuthal asymmetries are extracted in bins of $x$, $z = z_1 + z_2$ and the invariant mass $M_{\text{inv}}$, including a correction per kinematic bin regarding the beam polarization, the target polarization, the dilution of the target, as well as for respective depolarization factors.

\section{Results}

\begin{figure}
\includegraphics[width=0.49\textwidth]{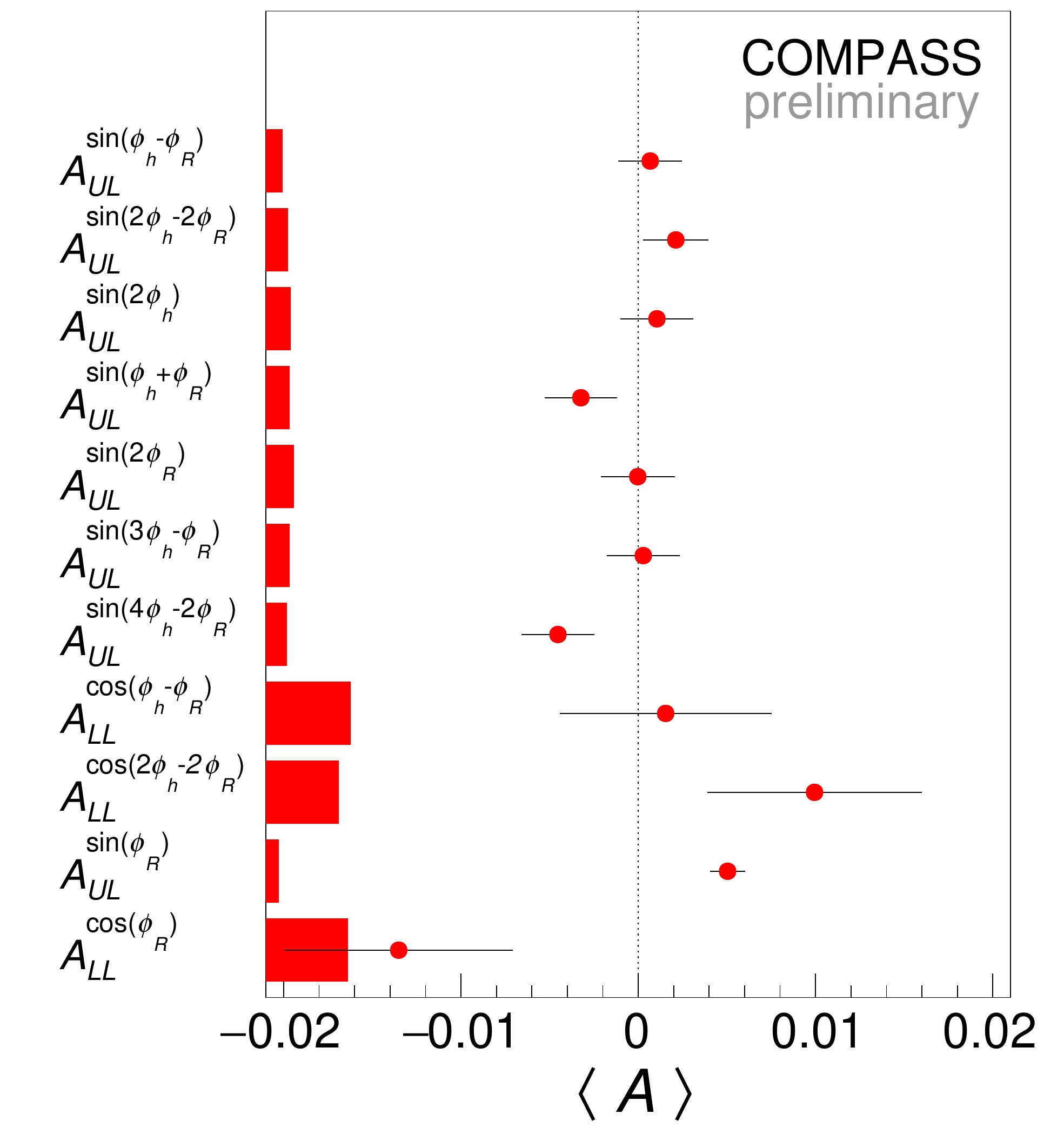}
\caption{Measured integrated azimuthal asymmetries arising in the di-hadron cross-section up to subleading twist, considering scattering off longitudinally polarized protons. Shown are the mean values when integrating over the entire kinematic range. The upper nine values correspond to asymmetries arising in a TMD approach at leading twist while the last two refer to the asymmetries at subleading twist in a collinear approach.}
\label{fig:A_Mean}
\end{figure}

\begin{figure*}
\includegraphics[width=.85\textwidth]{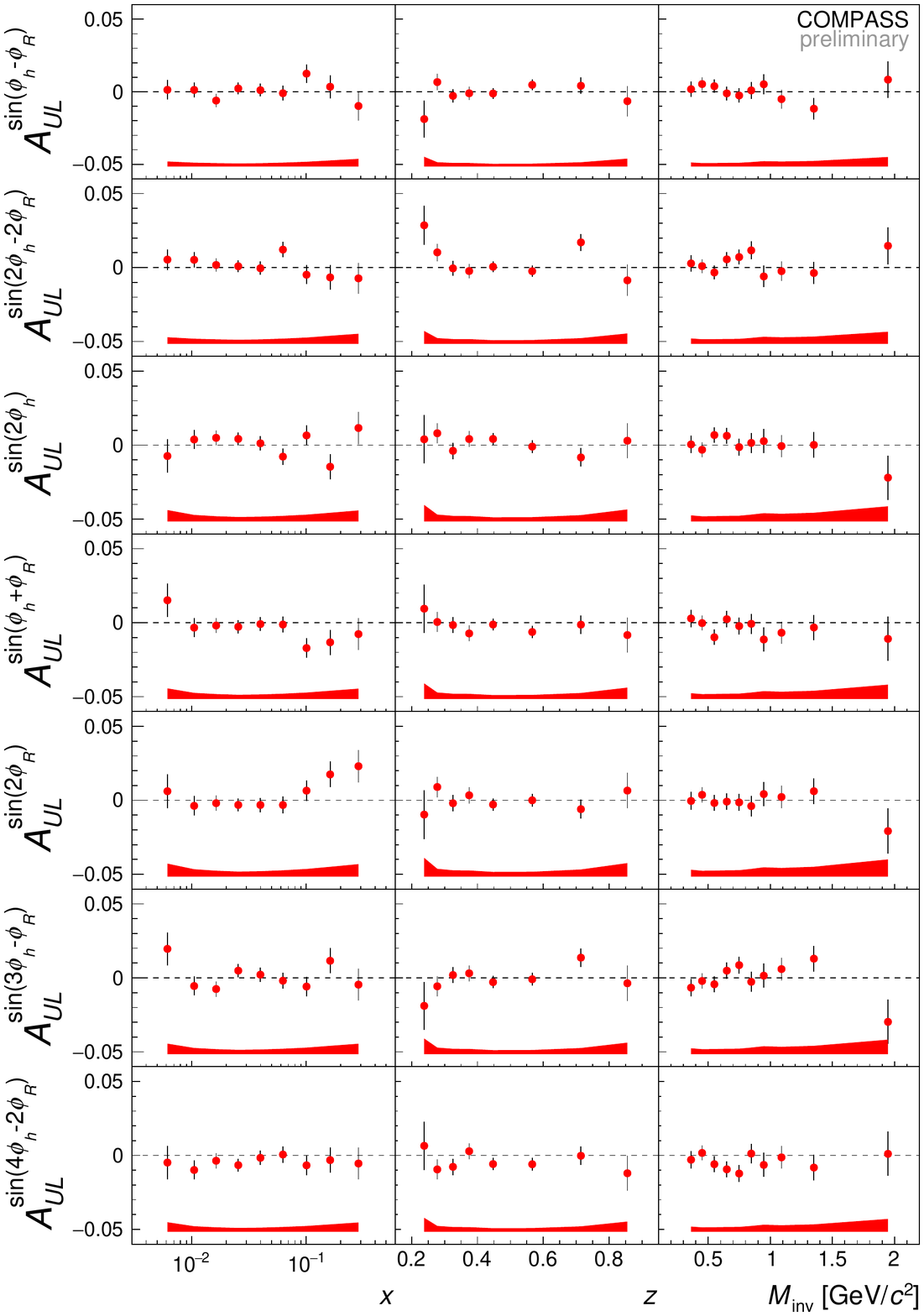}
\caption{Results for azimuthal SSAs arising in the TMD di-hadron cross-section at leading twist, considering scattering off longitudinally polarized protons.}
\label{fig:A_UL_T2}
\end{figure*}

\begin{figure*}
\includegraphics[width=.85\textwidth]{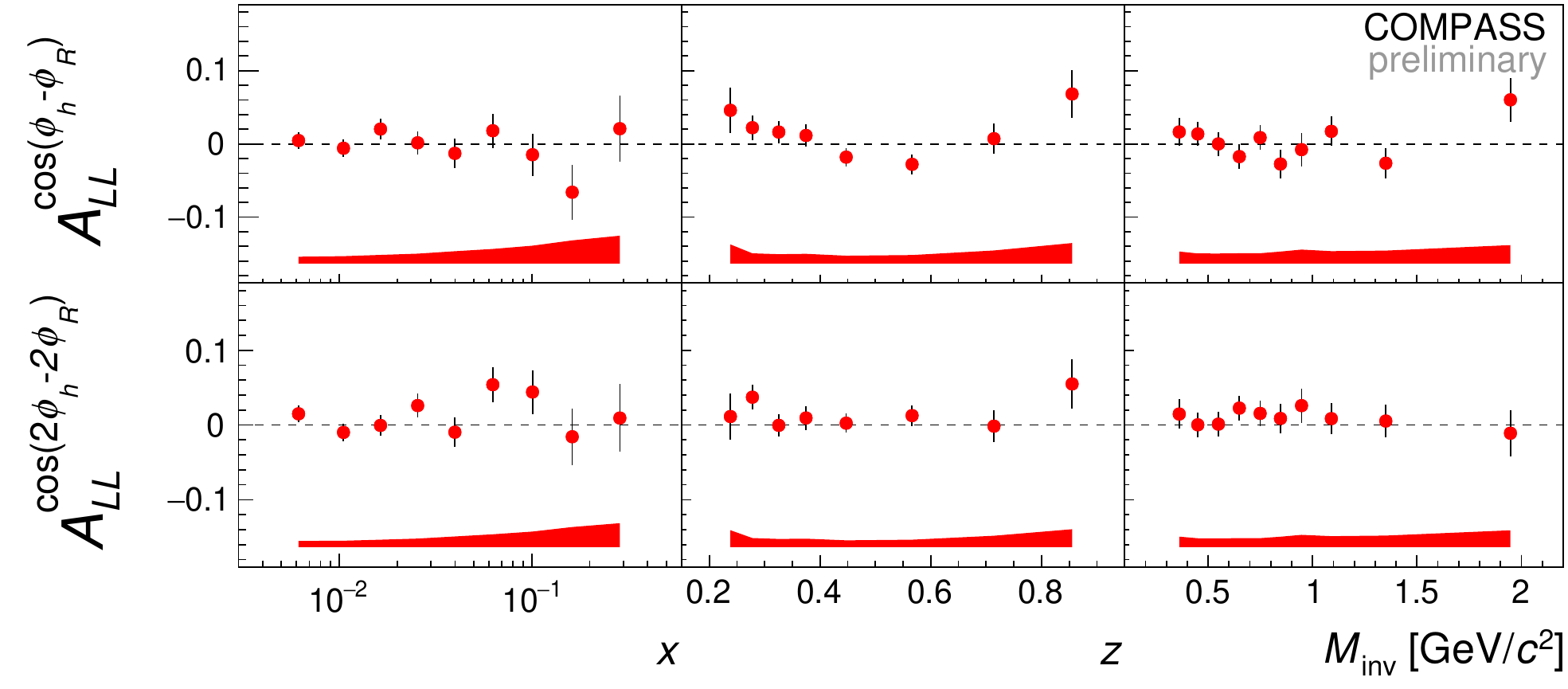}
\caption{Results for azimuthal DSAs arising in the TMD di-hadron cross-section at leading twist, considering scattering off longitudinally polarized protons.}
\label{fig:A_LL_T2}
\end{figure*}

\begin{figure*}
\includegraphics[width=.85\textwidth]{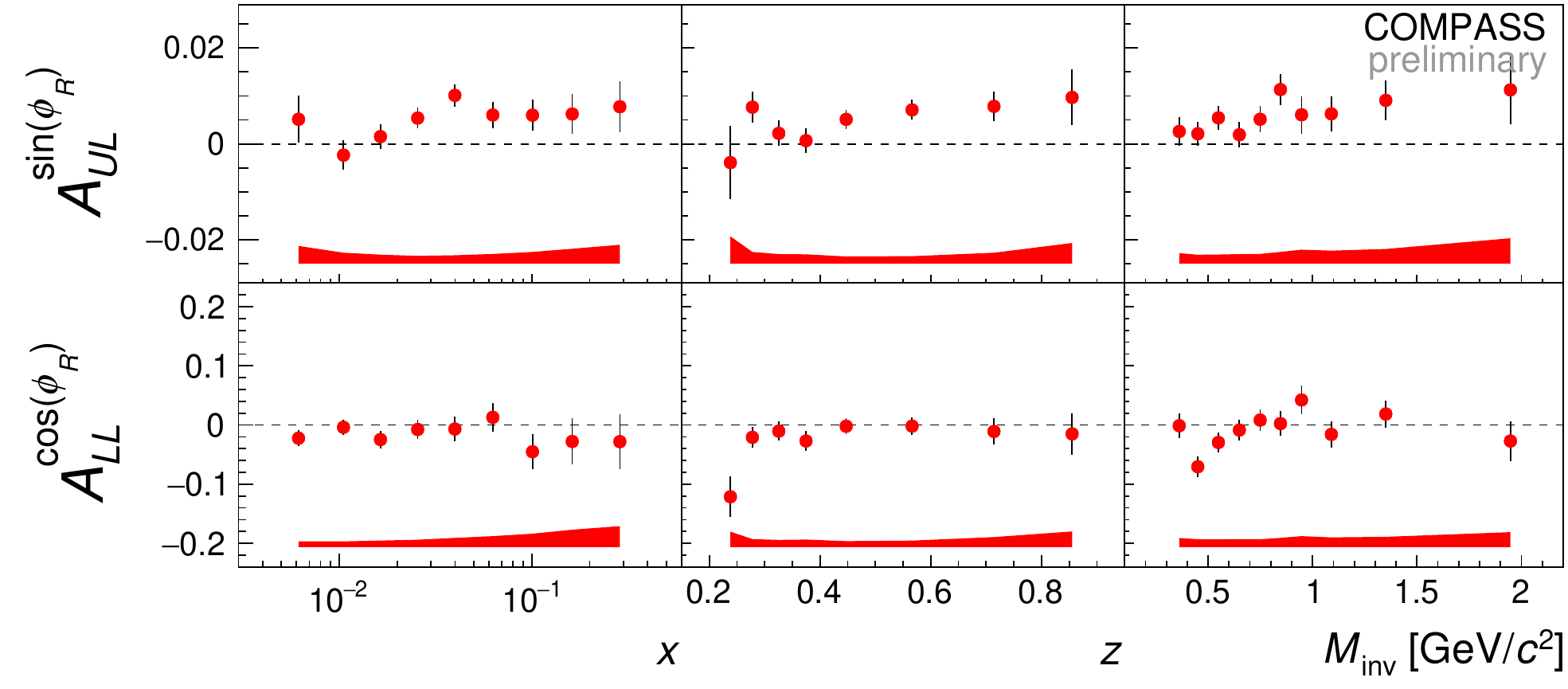}
\caption{Results for azimuthal asymmetries arising in the collinear di-hadron cross-section at subleading twist, considering scattering off longitudinally polarized protons.}
\label{fig:A_T3}
\end{figure*}

Our results for the asymmetries arising at leading twist are shown in Fig. \ref{fig:A_UL_T2} and Fig. \ref{fig:A_LL_T2}, where the statistical errors are represented by the error bars and the systematic uncertainties are indicated by color bands on the bottom of each plot. No eminent kinematic dependence is observed on any of the considered variables. The asymmetries are found to be quite narrowly distributed around zero over the entire kinematic ranges. \\
\par
Fig. \ref{fig:A_T3} shows our results for the two asymmetries at subleading twist. The single spin asymmetry $A_{UL}^{\sin(\phi_R)}$ is found to be clearly positive within experimental precision, averaging
\begin{equation}
A_{UL}^{\sin(\phi_R)} = \hspace{0.5cm}0.0050 \pm 0.0010(\text{stat}) \pm 0.0007(\text{sys}).
\end{equation}
This measurement confirms non-zero results from CLAS, measured in the high $x$-region. As already motivated in Sec. \ref{sec:Theo} the presented results can serve to access the still unknown PDF $h_L(x)$.\\
\par
The double spin asymmetry $A_{LL}^{\cos(\phi_R)}$ was found to average
\begin{equation}
A_{LL}^{\cos(\phi_R)} = -0.0135 \pm 0.0064(\text{stat}) \pm 0.0046(\text{sys}).
\end{equation}
The fact, that this asymmetry is found to be small within the experimental precision could consequently corroborate the Wandzura-Wilzcek assumption of negligible quark-gluon correlations on the fragmentation side, here encoded in the pure twist-3 interference FF $\tilde D^{\angle}$. The whole set of mean asymmetries measured within this work is shown in Fig. \ref{fig:A_Mean}.\\
\par
Summarizing the presented analysis, the obtained results provide an abundance of new information on spin related mechanisms inside hadrons and fragmentation processes. They can serve as valuable input for global analyses of PDFs and FFs as well as for the validation of theoretical model approaches.\\

\bibliography{Bib_Sirtl.bib}

\end{document}